\documentclass{aastex}
\usepackage{emulateapj5}
\usepackage{epsfig}

\newcommand{\etal}{\mbox{et al.}}
\newcommand{\ksxrb}{\mbox{KS 1731$-$260}}
\newcommand{\aqlxone}{\mbox{Aql X-1}}
\newcommand{\mxbgc}{\mbox{MXB 1743$-$29}}
\newcommand{\sixb}{\mbox{4U 1636$-$536}}
\newcommand{\sevenb}{\mbox{4U 1702$-$429}}
\newcommand{\slowb}{\mbox{4U 1728$-$34}}
\newcommand{\novab}{\mbox{4U 1608$-$522}}
\newcommand{\degdip}{\mbox{4U 1916$-$053}}
\newcommand{\mxbecl}{\mbox{MXB 1659$-$298}}

\makeatletter
\newenvironment{inlinefigure}{%
\def\@captype{figure}%
\noindent\begin{minipage}{0.999\linewidth}\begin{center}}
{\end{center}\end{minipage}\smallskip}
\makeatother

\slugcomment{To appear in \apj, v580, No.\ 2}
\begin{document}

\shortauthors{Muno \etal}
\shorttitle{Stability of Burst Oscillations}

\title{The Frequency Stability of Millisecond Oscillations in 
Thermonuclear X-Ray Bursts}
\author{Michael P. Muno, Deepto Chakrabarty\altaffilmark{1}, 
Duncan K. Galloway}
\affil{Department of Physics and Center for Space Research, 
       Massachusetts Institute of Technology, Cambridge, MA 02139}
\email{muno,deepto,duncan@space.mit.edu}
\and
\author{Dimitrios Psaltis}
\affil{School of Natural Sciences, Institute for Advanced Study,
       Princeton, NJ 08540}
\email{dpsaltis@ias.edu}
\altaffiltext{1}{Alfred P. Sloan Research Fellow}

\begin{abstract}
We analyze the frequency evolution of millisecond oscillations observed 
during type~I X-ray bursts with the {\it Rossi X-ray Timing Explorer}
in order to establish the stability of the mechanism underlying the 
oscillations. Our sample contains 68 pulse trains 
detected in a search of 159 bursts from 8 accreting neutron 
stars. As a first step, we confirm that the oscillations 
usually drift upward in frequency by about 1\% toward an apparent saturation 
frequency. Previously noted anomalies, such as drifts toward lower frequencies
as the oscillations disappear (``spin-down'' episodes) and instances of two 
signals present simultaneously at frequencies separated by a few Hz, occur 
in 5\% of oscillations. Having verified the generally accepted 
description of burst oscillations, we proceed to study the coherence of the 
oscillations during individual bursts, 
and the dispersion in the asymptotic frequencies in bursts observed over 
five years. On short time scales, 
we find that 30\% of the oscillation trains do not appear to evolve smoothly 
in phase. This suggests either that two signals are present simultaneously 
with a frequency difference too small to resolve ($\lesssim 1$~Hz), that
the frequency evolution is discontinuous, or that discrete phase jumps occur.
On time scales of years, the maximum 
frequencies of the oscillations exhibit fractional dispersions of 
$\Delta\nu_{\rm max}/\langle\nu_{\rm max}\rangle \la 4\times 10^{-3}$.
In the case of \sixb, this dispersion is uncorrelated with the known orbital
phase, which indicates that a mechanism besides 
orbital Doppler shifts prevents the oscillations from appearing perfectly 
stable. 
In the course of this analysis, we also search for connections between the 
properties of the oscillations and the underlying bursts. We find that 
the magnitudes of the observed frequency drifts are largest when the 
oscillations are first observed at the start of the burst, which suggests 
that their evolution begins when the burst is ignited. We also find that 
radius expansion appears to temporarily interrupt the oscillation 
trains. 
We interpret these results under the assumption that the 
oscillations originate from anisotropies in the emission from 
the surfaces of these rotating neutron stars. 

\end{abstract}

\section{Introduction}
Millisecond oscillations have been observed during thermonuclear X-ray bursts
from nine neutron stars in low mass X-ray binaries
\citep[LMXBs; see][for a review]{str01}. 
The bursts are triggered by the unstable 
nuclear burning of accreted material on the neutron star's surface
\citep[see][for a review]{lvt93}. It therefore has long been expected that 
anisotropies in the burning should produce pulsations at the stellar spin 
frequency \citep[e.g][]{bil95, str96}. Indeed, the general characteristics 
of the observed oscillations suggest that they originate from a
brightness anisotropy 
on the stellar surface. First, the oscillations are highly coherent towards 
the end of the burst \citep{str96, sm99}. Second, the oscillations are 
frequently observed in the rise of bursts, when there is spectral evidence 
for a growing burning region \citep{szs97}. Third, the frequencies of
these oscillations are remarkably similar in bursts separated by several
years \citep{str98, mun00}, which suggests that they originate from 
a stable clock. 

If the burst oscillation frequency,
$\nu_{\rm burst}$, is indeed the spin frequency of the neutron star, 
the oscillations are key to understanding several aspects of these systems. 
Pairs of kilohertz 
quasi-periodic oscillations (kHz QPOs) are observed in the persistent emission 
from many neutron star LMXBs. Their frequency difference 
$\Delta\nu_{\rm kHz}$ remains
comparable to $0.5$ or $1\times\nu_{\rm burst}$, even though the frequencies
of the kHz QPOs vary by more than a factor of 2 
\citep[see][for a review]{vdk00}. It has been suggested 
that these QPOs result from signals at a Keplerian frequency in the
disk and a beat between that frequency and the spin of the neutron star
\citep{str96, mlp98}.\footnote{Alternative models for the kHz QPOs cannot 
account naturally for the similarity between $\nu_{\rm burst}$ and 
$\nu_{\rm kHz}$ \citep[e.g][]{psa01}.} 
If this is the case, we can infer the rotation rate of the neutron stars in 
more than 20 systems \citep{vdk00}. The inferred spin frequencies cluster
around $300$~Hz. This confirms the suggestion that these neutron 
star LMXBs are the progenitors of recycled millisecond radio pulsars 
\citep{alp82,rs82}. However, it also implies that some 
mechanism limits the frequencies to which these stars can be spun-up by 
accretion, such as a correlation between the surface magnetic field strength 
and X-ray luminosity \citep{wz97} or gravitational radiation from a 
quadrupole in the mass distribution of the star \citep{bil98a}. 

A closer examination of these burst oscillations reveals an even 
more complex picture \citep[see also][for a detailed discussion]{psa01}. 
First, the coherence of the oscillations is cast into
some doubt by the detection of a sudden 0.25 cycle phase shift in one 
oscillation out of seven studied by \citet{str01}. Second, oscillations are 
not only observed in the rise of thermonuclear bursts, but also far into 
the decay \citep[e.g.][]{smb97}, even though by this time the 
fuel over the entire 
surface of the neutron star is expected to have been consumed by the 
burning front leaving no observable brightness anisotropies 
\citep[e.g.][]{fw82}. Third, while the spin of the neutron star 
must be constant during the course of a burst, the frequency of the 
oscillations is usually observed to increase and saturate at some maximum 
frequency \citep{str97}. It is this ``asymptotic'' frequency that appears 
stable in bursts separated by several years. The frequency drift is 
thought to originate from one of several mechanisms that might carry
anisotropies in a retrograde motion about the neutron star, resulting
in an observed frequency slightly lower than that of the star's spin
(Strohmayer et al. 1997; Cumming \& Bildsten 2000; Spitkovsky, Levin, 
and Ushomirsky 2002; Heyl 2002).
Finally, in a handful of bursts there are episodes of drifts to lower 
frequency (``spin-down'') in the tails of the bursts 
\citep{str99, mil00, mun00} and strong detections of 
signals at second frequencies within a few Hz of the primary signal 
\citep{mil00,gal00}.

If anything is clear, it is that more observational clues are required to
understand burst oscillations. In this paper, we address the how the
frequency of these oscillations evolves as a function of time, in order
to establish the coherence of the underlying mechanism.
In particular, we employ standard pulse phase 
analysis techniques adapted from pulsar timing studies \citep{mt77}.

\section{Observations and Data Analysis}

Our analysis used observations with the Proportional Counter Array
(PCA; Jahoda et al. 1996) on the {\em Rossi X-Ray Timing Explorer}
(RXTE).  PCA consists of five identical gas-filled proportional
counter units with a total effective area of 6000~cm$^2$ and
sensitivity in the 2.5--60~keV range.  The detector is capable of
recording photons with microsecond time resolution and 256-channel
energy resolution.   The data were recorded
in a wide variety of data modes with different time and energy
resolutions, depending upon the details of the original proposed
programs and the available telemetry bandwidth.   For all of the analysis
presented here, we converted the TT photon arrival times at the
spacecraft to Barycentric Dynamical Time (TDB)

\begin{inlinefigure}
\centerline{\epsfig{file=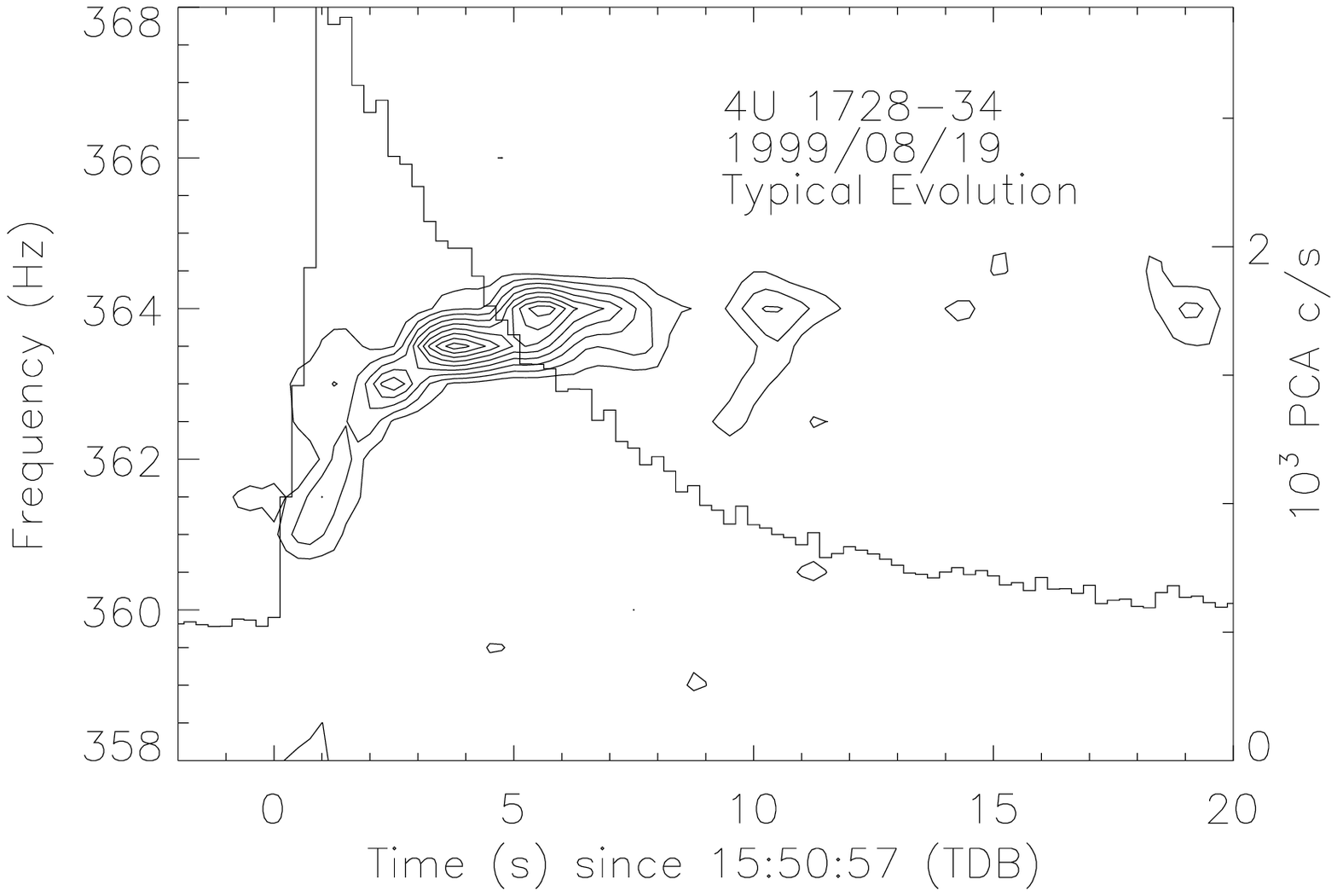,width=0.9\linewidth}}
\caption{A dynamic power spectrum illustrating the typical frequency
evolution of a burst oscillation. Contours of power as a function of 
frequency and time were generated from power spectra of 2 s intervals
computed every 0.25 s. A Welch function was used to taper the data 
to reduce sidebands in the power spectrum due to its finite length
\citep{pre92}. The contour levels are at powers of $0.02$ in single-trial
probability starting at a 
chance occurrence of $0.02$. The PCA count rate is plotted referenced to 
the right axis.
\label{fig:typical}}
\end{inlinefigure}

\noindent
at the solar system
barycenter, using the Jet Propulsion Laboratory DE-200 solar system 
ephemeris (Standish et al. 1992). 

We searched the entire {\em RXTE} public data archive for
X-ray bursts from 8 neutron stars\footnote{Oscillations in bursts
associated with MXB~1743$-$29 have been observed during observations
of the bursting pulsar GRO~J1744$-$28.  A search for these bursts was
not part of the analysis presented here.} that are known to exhibit
burst oscillations (see Table 1 and Muno et al. 2001).  As of
September 2001, we have identified a total of 159 X-ray bursts from
these 8 sources.  Each of these bursts was then searched for millisecond
oscillations as described below.

For our analysis, we used data containing a single energy
channel (2.5--60 keV) and $2^{-13}$~s (122 $\mu$s) resolution.  We note
that the use of more restricted energy bands can sometimes allow oscillation
trains to be detected for a second or two longer, since the oscillation
amplitudes are often stronger at higher energies (e.g., Muno et al. 2000;
Giles et al. 2002).  However, in many cases, the data modes with higher
energy resolution contained data gaps due to the limited size of the
satellite memory buffers, making it difficult to produce a coherent
phase model spanning the entire oscillation train (see Section~2.2).  
Data modes specifically designed to start recording during a burst
(``burst catcher'' modes) were analyzed in combination with the other
modes in order to fill data gaps, but the burst catcher modes with high 
time resolution only recorded in a single 2.5--60~keV energy channel.

\subsection{Fourier Timing}

We produced Fourier power spectra of 1 s intervals of data for the first 
16 s of each burst, and searched for signals within a frequency range of 
$\pm5$~Hz from the frequencies listed in Table~\ref{sources}. 
Any oscillation with a single trial probability of less
than $2\times10^{-5}$ that it is due to noise (a power of 
$P > 21.6$ according to the normalization of Leahy et al.\ 1983) was 
considered a detection. This corresponds to a 1\% probability that a 
noise signal will be that strong given a search of an entire burst, so that
1--2 noise signals may appear significant from the search of our entire
sample of bursts. We detected 68 oscillation trains
 in this manner 
(Table~\ref{sources}). We also recorded the frequencies of each signal 
detected. The frequencies in Table~\ref{sources} were determined
recursively, by computing the 
median value (rounded to the nearest Hz)
from the signals detected until that value did not change.
The range of detected frequencies is reported in Section~3.1.

To qualitatively examine the frequency evolution, we computed power
spectra of 2 s intervals of data every 0.25 s, and plotted contours 
of power as a function of both time and frequency. We tapered the data
with a Welch function \citep{pre92} before computing each power spectrum, 
in order to suppress the power in sidebands due to the window function. A 
typical example of such a dynamic power spectrum is illustrated in 
Figure~\ref{fig:typical}. We use dynamic power spectra to characterize
the oscillations in Sections~3.2--3.3.
 
\subsection{Phase Connection}

In order to examine in detail whether the oscillations are coherent and drift 
to stable asymptotic frequencies, we 
developed frequency models using the phase connection method described 
in \citet{mun00}. We modeled all oscillations that lasted continuously 
for more than 2 s using the $2^{-13}$~s data (Table~\ref{sources}). 
A summary of the procedure is as follows. 
We fold the data in 0.25--0.5 s intervals
 according to a trial frequency 
model (which is the derivative of a phase model), and compute a Fourier
transform of the resulting profile. If the power $P$ in the 
fundamental frequency of the pulsation\footnote{We have confirmed that all 
of the profiles were consistent with sinusoidal signals \citep{mun00,mun01},
justifying the use of only the power at the fundamental frequency.}
has less than a 10\% chance of 
occurring randomly due to noise ($P > 4.6$ in the normalization of 
Leahy et al.\ 1983), we measure its phase by a linear least 
squares fit of a sinusoid to the profile. The measured phase residuals
$\phi_{\rm obs}(t_i)$ are then 
related to the predicted phases $\phi_{\rm model}(t_i)$ by
\begin{equation}\label{eq:dph}
\Delta\phi(t_i) \equiv \phi_{\rm obs}(t_i) - \phi_{\rm model}(t_i).
\end{equation}
In the above convention, an increasing phase residual indicates that the
instantaneous trial frequency ($d\phi_{\rm model}/dt_i$) is too low.

The phase residuals $\Delta\phi(t_i)$ are then modeled via a least squares 
minimization, using a $\chi^2$ statistic calculated from the phase 
residuals: 
\begin{equation}\label{eq:chisq}
\chi^2 = \sum {{(\Delta\phi(t_i))^2} \over {(\sigma_{\Delta\phi}})^2}.
\end{equation}
Here, $\sigma_{\Delta\phi}$ is the uncertainty on the phase measured
from the folded profile, which can easily be determined from the diagonal 
values of the covariance matrix in the linear least squares fit \citep{pre92}. 
We have confirmed that these uncertainties are accurate via Monte Carlo 
simulations, in which we ({\it i}) perturbed the counts in each bin of 
the profile within the range expected from Poisson counting noise and 
({\it ii}) recorded the uncertainty as the standard deviation of phase 
measurements of 100 perturbed profiles. We also verified the values
for $\sigma_{\Delta\phi}$ by visual inspection.

The phase connection technique provides a correction $d(\Delta\phi)/dt$ to 
the original frequency model $\nu_{\rm model}$, so that the best-fit
frequency is 
\begin{equation}
\nu(t) = \nu_{\rm model}(t) + {{d}\over{dt}}(\Delta\phi)
\end{equation}
This procedure can be iterated using the new frequency model, if it 
is composed of linear functions.

The advantages of this method are that ({\it i}) the form 
of the observed phase residuals naturally suggests the best model for the
frequency evolution and provides a quantitative test of how well  
the trial phase evolution matches the data, ({\it ii})
there exist standard techniques to calculate the uncertainties on 
the model from the least squares fit to the phases, and ({\it iii})
the method utilizes the 
phase information that the power spectrum discards, which allows
much finer frequency resolution on short time scales. In contrast, 
the $Z^2$ method of \citet{sm99} distinguishes between models based upon 
the amount of power in the folded profile, but provides little information 
about how the frequency of an oscillation deviates from the
model assumed. The $Z^2$ statistic is still very useful for examining the 
amplitude and phase evolution of the oscillations 
on short (less than 0.1 s) time scales after the frequency evolution is 
modeled. We apply this technique and use the resulting models in 
Sections~3.4--3.7.

\subsection{Energy Spectra}

Finally, we produced energy spectra in 0.25 s intervals from each burst
using available combinations of data modes that provide at least 32 energy
channels. 
The detector response was estimated using PCARSP in FTOOLS version 5.1
(see http://heasarc.gsfc.nasa.gov/lheasoft/).
We subtracted spectra from 15 s of emission prior to the burst 
to account for background, and fit each spectrum between 2.5--20 keV 
with a model consisting of a blackbody multiplied by a constant interstellar 
absorption factor. The absorbing column for each burst was taken to be the 
mean value from fits to each 0.25 s interval assuming a variable absorption. 
The model provides a color temperature ($T_{\rm col}$) and
a normalization proportional to the square of the apparent radius 
of the burst emission surface ($R_{\rm col}$), and allows us to estimate the
bolometric flux as a function of time. We compare these spectra 
to the properties of the oscillations in Sections~3.3 and 3.6.

\section{Results}

\subsection{Frequency Range of Oscillations}

\begin{figure*}[t]
\centerline{\epsfig{file=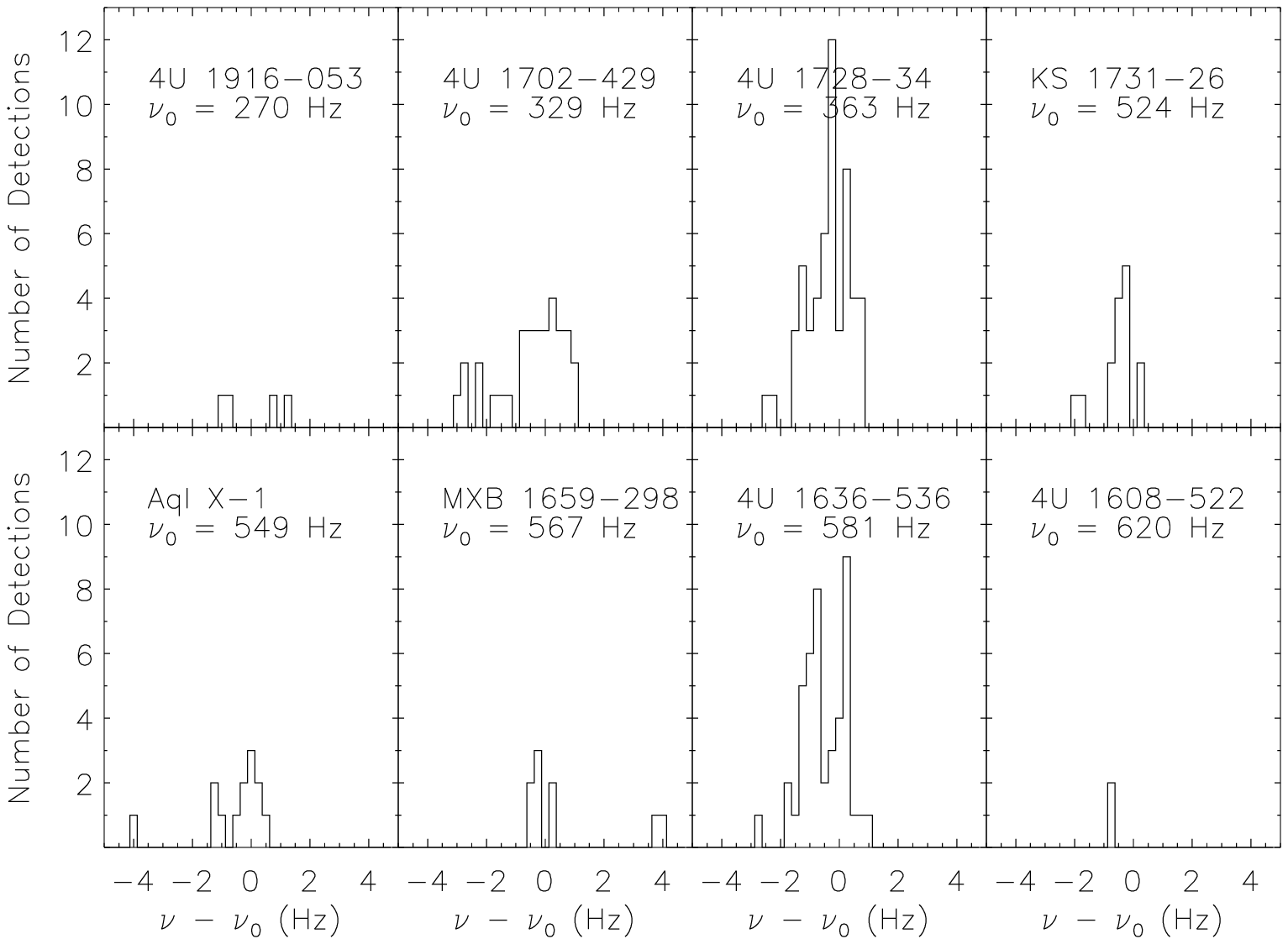,width=0.9\linewidth}}
\caption{Histograms of the number of oscillations detected at each 
frequency within a 10 Hz range of the mean indicated during the first 
16 s of each burst, separated by source. Detections are 
defined as having a single-trial probability of less than $2\times10^{-5}$
that they are due to noise. The histograms are fairly narrowly 
distributed 
with an absolute width of about 4 Hz. Note that the signal at 571~Hz
in \mxbecl\ has a chance probability of 3$\times10^{-4}$ given that
we searched 159 bursts from all sources to produce the histograms.
\label{hist}}
\end{figure*}

In order to establish the range of frequencies over which the oscillations 
are observed, and in particular whether there are strict upper limits
to their frequencies, we produced a histogram of the distribution of 
the frequencies at which signals were detected in non-overlapping 1 s intervals
during the bursts from 
each source (Figure~\ref{hist}). For the majority of sources, 
the histograms are narrowly distributed with a width of 2-4~Hz, significantly
smaller than the 10~Hz search window. This 
corresponds to frequency drifts ranging from 0.4\% (\ksxrb) to
1.2\% (\sevenb).

The distribution from \sixb\ appears to be double-peaked, possibly 
owing to two effects. First, oscillations are usually observed from
this source only in the rise and tail, and the frequency gap between 
the two peaks could result from the fact that we do not observe the 
intermediate frequencies as the oscillations evolve (see Section~3.2). 
However, two oscillations are present simultaneously in several bursts 
from this source (see

\noindent
Section~3.3 and Miller 2000), which could indicate 
there are indeed two fundamental frequencies.
 
One detected oscillation is unusual given the distributions in 
Figure~\ref{hist}, and has already been noted by Wijnands, Strohmayer, \&
Franco (2000). It 
is detected in \mxbecl\ about 4 Hz above the next highest frequency,
with a probability of $4\times10^{-4}$ that it is due to noise given all 
of the trial frequencies searched. The oscillation occurs at 571~Hz in 
a single interval about 15 s
into the tail of a burst, while the oscillation detected earlier
in the burst seemed 
to disappear about 5 s into the burst after reaching an asymptotic frequency 
of 567 Hz. If this

 signal is a burst oscillation, it calls into question 
whether we are observing frequency saturation at all. 

\subsection{Instances of Unusual Frequency Evolution}

We also searched the dynamic power spectra to confirm that the oscillations
in our sample drift upward in frequency during the course of the 
burst, saturating at an approximately constant frequency late in the tail.
This trend occurs in most bursts, and is illustrated in 
Figure~\ref{fig:typical}. However, there are a few exceptions
that have been noted in the 
literature, such as 
({\it i}) clear evidence for a frequency {\it decrease} several 
seconds after the burst is first detected (spin-down) and ({\it ii}) signals 
present at two frequencies simultaneously. 
Spin-down episodes and simultaneous signals are not common.

We find only 3 clear examples of spin-down
episodes during bursts, which we display in Figure~\ref{sdown}. 
Two of these have been reported previously, from \sixb\ \citep{str99} and 
\ksxrb\ \citep{mun00}, and we find one additional example from \slowb. 
In all cases, the frequency decrease is on order 1~Hz. There is no clear 
pattern as to when in the burst the spin-down 
occurs. In \ksxrb, the event occurs in the peak of the burst, 
while in \slowb\ and \sixb\ the spin-down occurs as the burst decays.

We also find two
clear examples of bursts in which two signals are detected simultaneously
at frequencies separated by about 1~Hz. The most significant example 
has already been reported by \citet{mil00} from \sixb, and we find a 
second example in \sevenb. Both are displayed in 
Figure~\ref{fig:twofreq}. The second oscillation in \sixb\ occurs between
2--3 s into the burst, and has a chance probability $< 3 \times 10^{-10}$
of occurring randomly due to noise \citep[see also][]{mil00}. 
The second 
signal from \sevenb\ occurs 11 s into the burst, 
and is only marginally significant, with a single-trial chance 
probability of $5 \times 10^{-5}$. Less significant 
examples, with chance probabilities of $\sim 0.01$, have also been reported
in 

\begin{inlinefigure}
\centerline{\epsfig{file=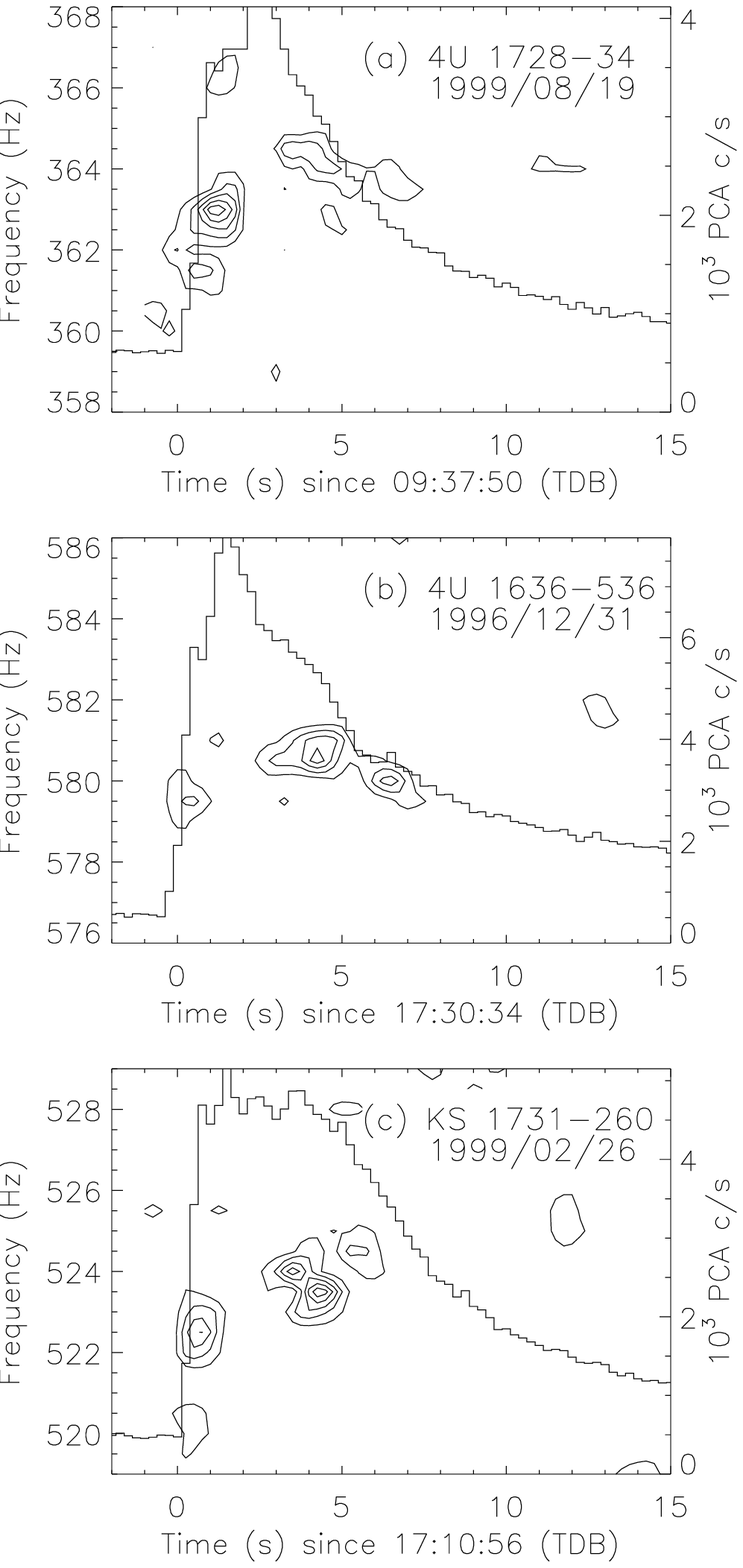,width=0.75\linewidth}}
\caption{Same as Figure~1, for the 3 oscillations that exhibited 
spin-down 
episodes out of our sample of 68 pulsations. The frequency drift is on
order 1~Hz in each case. There appears to be no trend as to when in 
the burst spin-down occurs.
\label{sdown}}
\end{inlinefigure}

\noindent
two more bursts from \sixb\ by \citet{mil00}, and in one burst 
from \degdip\ by \citet{gal00}. In all cases, 
the secondary frequencies are 1--3~Hz lower than the main signal.

\subsection{How Radius Expansion Affects Oscillations}

\begin{figure*}[p]
\centerline{\epsfig{file=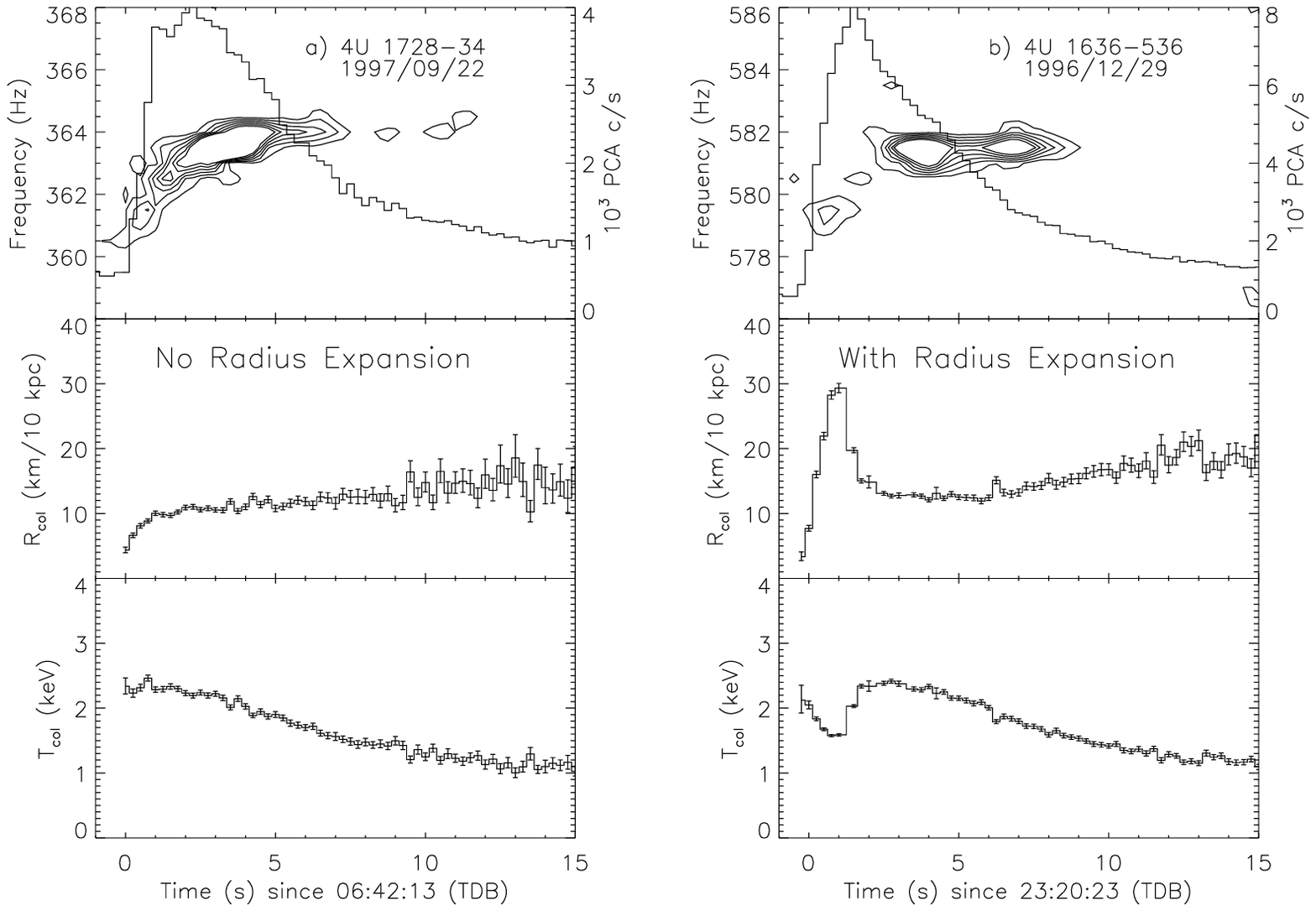,width=0.85\linewidth}}
\caption{Dynamic power spectra compared to parameters from blackbody 
fits to the energy spectra.
{\it Top Panels}: Same as Figure~1. {\it Center Panels}: The color 
radius derived from the spectral fits, as a function of time. 
{\it Bottom Panels}: The color temperature of the burst.  Uncertainties
are 1-$\sigma$. No radius expansion occurred during the burst on the 
left, and the oscillations were observed continuously for the first 7~s of
the burst. The radius expansion on the right appears to interrupt 
the oscillations. Note that the power spectra are of 2~s intervals
spaced every 0.25 s, while the energy spectra are computed in 
non-overlapping 0.25 s intervals. 
\label{fig:pre}
}
\end{figure*}

In many bursts, photospheric radius expansion is evident at the start
of the burst, during which $R_{\rm col}$ increases and $T_{\rm col}$
decreases such that the bolometric flux remains constant, presumably
at the Eddington limit (see Lewin et al. 1993). We plot examples of bursts
with and without radius expansion in Figure~\ref{fig:pre}.
We examined whether radius expansion affects the times in the burst 
when the oscillations are observed. We find that all 16 of the bursts in
which oscillation trains are observed continuously throughout 
the rises, peaks, and tails fail to exhibit radius expansion 
(e.g. Figure~\ref{fig:pre}a). Of the 51 bursts where the oscillation train 
is not observed during the peak, 40 exhibit radius expansion 
(e.g. Figure~\ref{fig:pre}b). 
One burst from \novab\ (1998 March 27 at 14:08:30 TDB) could not 

\begin{inlinefigure}
\centerline{\epsfig{file=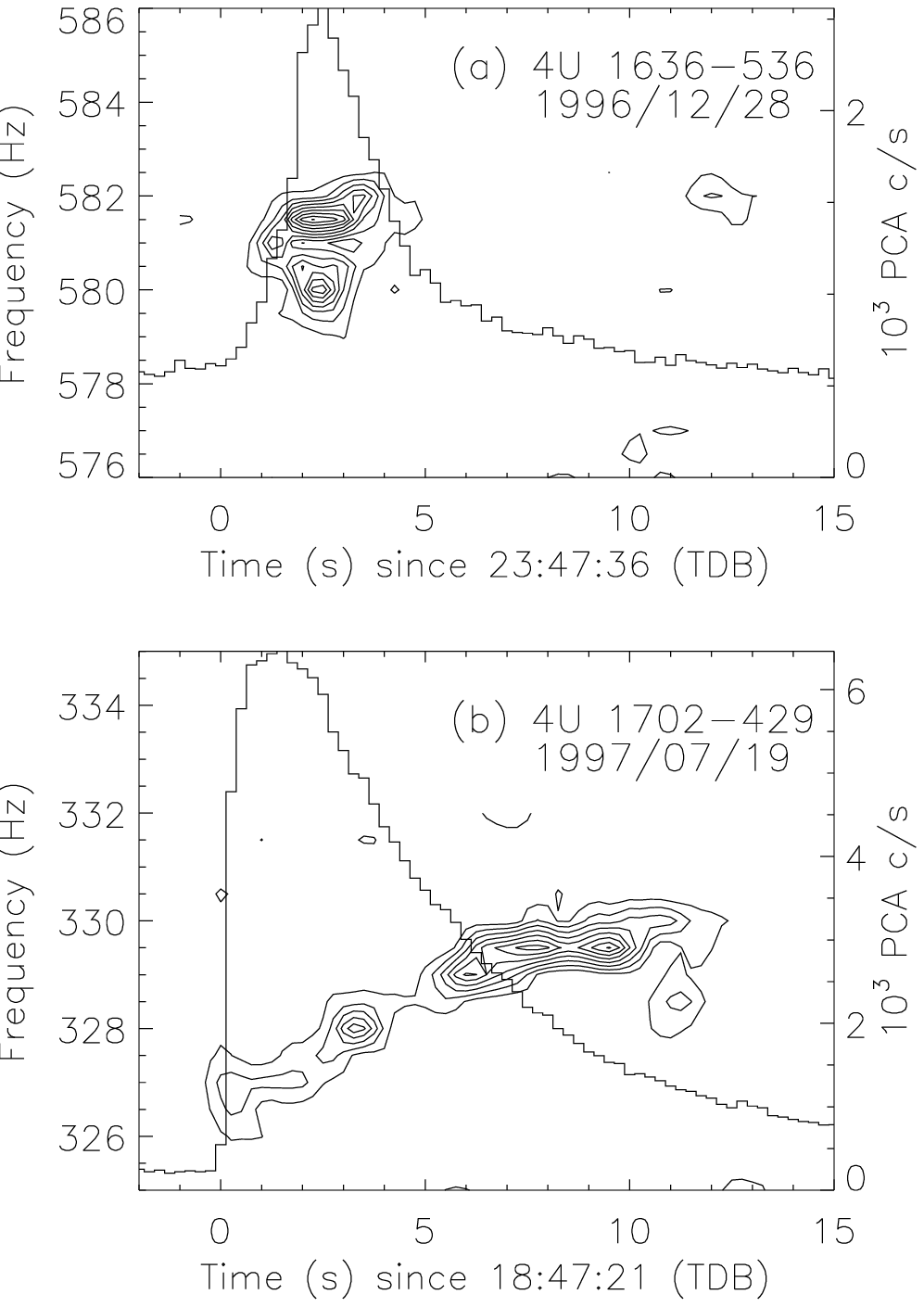,width=0.75\linewidth}}
\caption{Same as Figure~1, for oscillations that exhibited two 
simultaneous
signals separated in frequency by $\sim 1$~Hz. The secondary signal 
occurs at 2--3~s in panel a, with a chance probability that it is due to
noise of $3\times10^{-10}$. The second signal at 11~s in panel b has a 
chance probability of $5 \times 10^{-5}$. 
\label{fig:twofreq}
}
\end{inlinefigure}

\noindent
be examined, because there were gaps in the high time
 resolution data during
the peak. Since it also has been noticed previously that oscillations in 
\ksxrb\ \citep{smb97} and \mxbgc\ \citep{str97} appear only 
after radius expansion ends, we conclude that radius expansion
either obscures or interrupts the oscillations.


\subsection{Mathematical Form of the Frequency (Phase) Evolution}

We modeled the frequency evolution of the oscillations using the phase 
connection technique described in Section~2.2. We tried
both polynomial models (up to fifth order) and a saturating 
exponential such that the frequency evolution is of the form used by 
\citet{sm99}: 
\begin{equation}\label{eq:chirp}
\nu(t) = \nu_0 + \Delta\nu \exp(-t/\tau).
\end{equation}
Here, $\nu_0$ is the asymptotic frequency of the oscillation, 
$\Delta\nu$ is the frequency drift referenced to the start of the burst, and
$\tau$ is the time scale for the frequency drift. 
When modeling the data with polynomials, we iterated the fitting procedure 
until 
no improvement in the fits to the phase residuals was found.
The fits using the exponential model were executed on phase residuals 
found assuming a constant trial frequency, as the corrections must add
linearly to the trial phase model in order to use the iterative process. 
Table~\ref{bestmod} lists the number of oscillations best fit with 
each model, as determined from the
value of reduced $\chi^2$. If the reduced $\chi^2$ was greater than one, 
and if both the number of degrees of freedom and the value of $\chi^2$ 
decreased in
comparing one model to another, an $F$-test \citep{bev69} was used to
ensure that the decrease in $\chi^2$ had less than a 5\% chance of being 
random.

\begin{figure*}[t]
\centerline{\epsfig{file=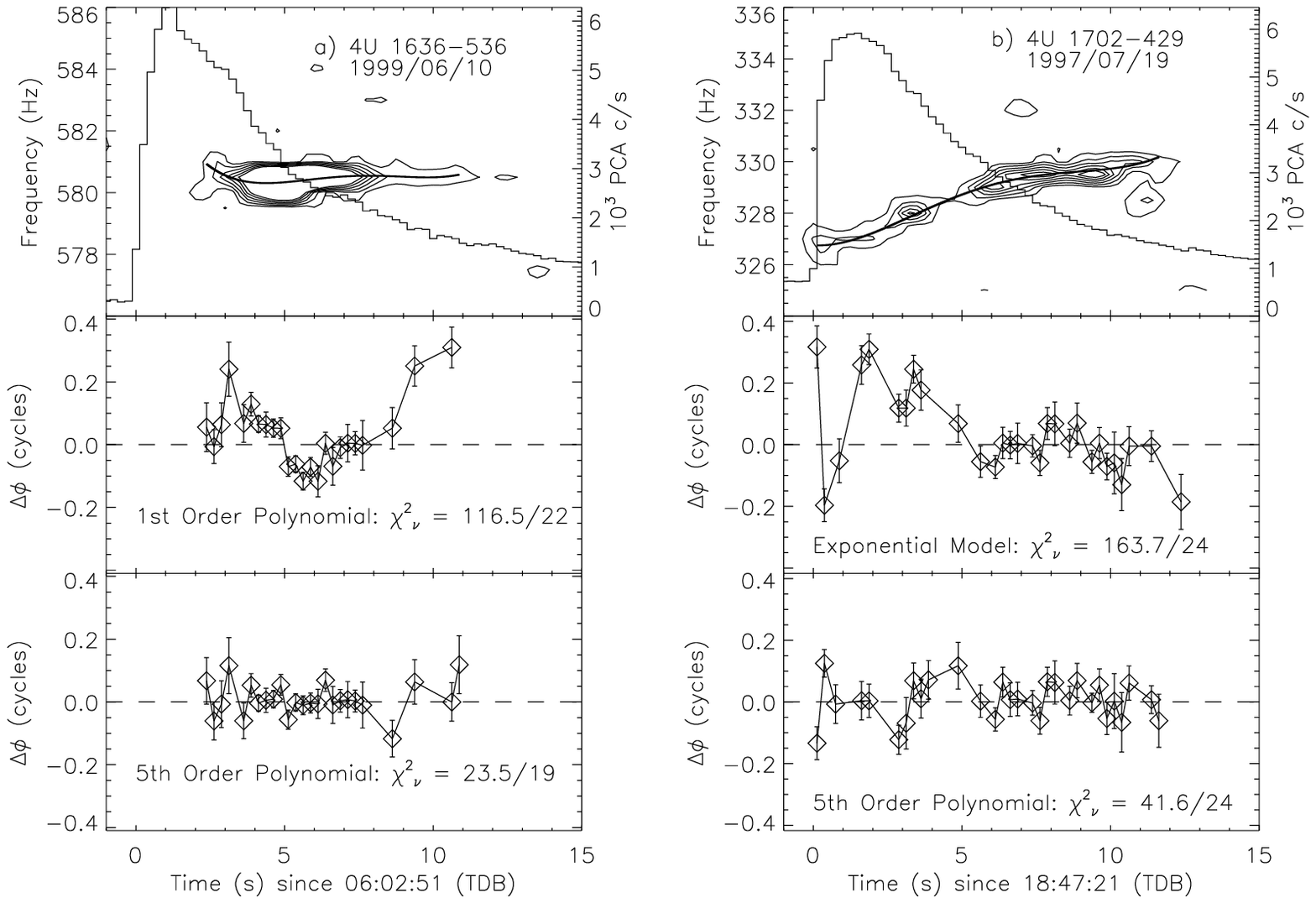,width=0.85\linewidth}}
\caption{The frequency and phase evolution of burst oscillations from 
(a) \sixb\ and (b) \sevenb. The figures demonstrate how 
the phase connection method allows one to distinguish between models 
for the frequency evolution. {\it Top panel:} Same as Figure~1. The solid 
line at the 
center of the contours is the best-fit frequency evolution. 
{\it Center and bottom panels:} The phase residuals
after folding a the data about the model for the frequency evolution 
indicated. The bottom panel is the best-fit, as indicated by the lower
value of $\chi^2_\nu$.
\label{consev}}
\end{figure*}

Figure~\ref{consev}a 
illustrates the importance of considering the phases in modeling the 
frequency evolution. Judging from power spectra alone, the  
frequency of this oscillation from \sixb\ appears roughly constant 
({\it top panel}). However, when one examines the 
phase residuals from a constant frequency model it is clear that 
additional 
evolution has occurred ({\it center panel}). An acceptable fit to the 
frequency evolution only can be achieved using a fifth order polynomial
({\it bottom panel}). Although some of this frequency evolution can 
be inferred from the power spectra alone \citep[see also][]{mil00}, by 
tracking the phase we are able to achieve high frequency resolution and to 
determine how coherent the oscillations are.

We modeled 59 oscillations from 6 different sources
(Table~\ref{sources}). The results are summarized in Table~\ref{bestmod}.
Thirty-seven oscillations required either an 
exponential model or a polynomial of third degree or higher. Exponential 
models are only favored over polynomials in only
6 out of 37 cases. The exponential models are formally inconsistent 
(greater than 99\% confidence) in 22 of the remaining 31 cases, while all 
such frequency models are inconsistent with the data in 12 out of 37 cases. 

The polynomial models generally reproduce the frequency 
evolution better than exponential models because there is slower frequency 
evolution at the start of the oscillation train, inflection in the evolution 
during the train, or a decrease in the oscillation frequency as the 
trains disappear \citep[see also][]{mil00}. Figure~\ref{consev} 
exhibits all of these variations in the frequency evolution. In addition 
to the upward drifts in frequency that saturate after a few seconds, 
the frequencies also tend to wander by about 0.1~Hz 
(see, e.g., the evolution in the top panel of Figure~\ref{consev}a).

We see no evidence for a rapid decrease in frequency at the start of the
oscillation that is comparable in magnitude to the later upward drift in
frequency. Such a frequency decrease could occur as the burning front
spreads around the surface of the neutron star \citep[e.g.,][]{str97,slu02}, 
but may be too rapid to observe using our analyses.

\subsection{Short-term Stability of the Burst Oscillations}

\begin{figure*}[t]
\centerline{\epsfig{file=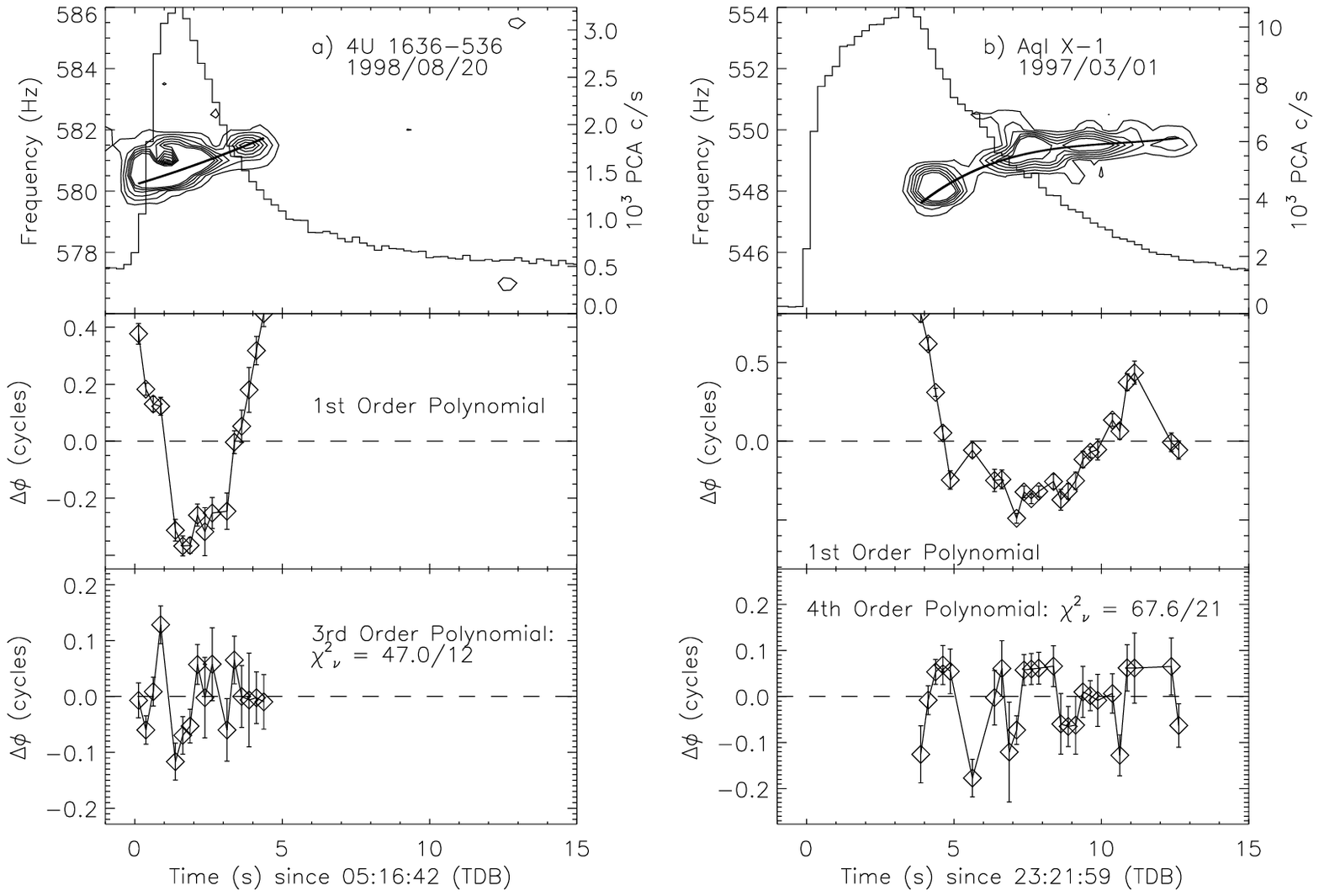,width=0.85\linewidth}}
\caption{The frequency and phase evolution of oscillations from two 
trains for which the best-fit model was statistically inconsistent with 
the data. The axes are as in Figure~6. 
\label{failed}}
\end{figure*}

In general, we find many more large values of $\chi^2$ than would be 
expected if the phase evolution were described by smooth, low-order 
polynomial and exponential functions. In particular, 30\% of the 
oscillation trains are inconsistent with smooth models with  
99\% confidence. This could indicate that the underlying mechanism is not 
perfectly stable. We tried higher-order polynomials in instances where 
our best-fit model is statistically unacceptable, but we find that 
we achieve only marginal improvements in $\chi^2$.
Higher-order polynomials can smooth residuals in the middle of the 
oscillation train, but tend to create larger residuals at the endpoints. 
Figure~\ref{failed} illustrates two examples of oscillation trains that 
we were unable to fit with 

\noindent
exponential or polynomial models. Piecewise 
smooth phase functions would provide better fits in some instances, such 
as between 2--4 s in Figure~\ref{failed}a, and 3--6 s in 
Figure~\ref{failed}b. However, the resulting discontinuities in the 
derivatives of the phases would imply that the frequencies of the oscillations
shift in less than 0.25 s.

Moreover, some of the residuals appear to be discrete changes in the 
phases of the oscillations. One possible explanation for the phase residuals 
in the first 1--2 s of Figure~\ref{failed}a is that two signals are 
present simultaneously within about 1 Hz, the resolution of the folding 
technique. Simultaneous signals with only slightly larger frequency 
separations have been seen in other bursts from \sixb\ \citep{mil00}.
In Figure~\ref{failed}b, there appear to be discrete 
jumps of about 0.1 cycle in the phase of the oscillation from 7--10 s into the
burst, similar to that observed by \citet{str01} during an oscillation in 
the rise of a burst from \sixb. We confirmed 
that the 
sudden changes in the phases of the oscillations are not due to 
variations in the pulse profile, which is always sinusoidal.
The presence of phase jumps suggests that the oscillations are not strictly
coherent.

\subsection{Magnitude and Time Scale of Frequency Drift}

Our models also allowed us to compare the frequency evolution
to other properties of the bursts. 
We first measured the observed change in frequency $\Delta\nu$ for the
continuous portion of each oscillation. We then measured the time scale 
$\tau$ that represents how long the oscillation took to evolve in frequency by 
$(1-e^{-1})\Delta\nu = 0.632\times\Delta\nu$ from the minimum 
observed frequency (compare Equation~2). 
As measures of the burst time scales, we have 
fit the decay in flux of each of these bursts with either one or two 
exponential functions ($\exp[-t/t_{d}]$). We also computed the peak 
bolometric flux and the fluence of each burst, using the exponential fit 
to determine the flux
contribution at late times. We find that neither 
$\Delta\nu$ nor $\tau$ are correlated with the above burst 
properties.

We have also compared $\Delta\nu$ and $\tau$ with the duration of the 
oscillations and the time at which they began.
We find that 
$\tau$ is not correlated with the start time or duration of the 
oscillations. Instead, $\tau$ is distributed uniformly 
between 0.75--5.5 s. We plot the frequency drift $\Delta\nu$ as a 
function of these 
parameters in Figure~\ref{tstart}. Oscillations that are first observed 
early in the burst 
evolve over a larger frequency range ({\it top panel}), but 
their frequency drift is not correlated with how long the oscillations 
last ({\it bottom panel}). This 
suggests that the evolution is initiated by the start of the burst.

Cummings \& Bildsten (2000) previously noticed that the fractional frequency 
drifts $\Delta\nu/\nu$
in sources with $\sim$600 Hz oscillations are a factor of two smaller 
than those in $\sim$300 Hz oscillations. We have indicated the data from
the faster ($\sim$600~Hz) 
oscillations in Figure~\ref{tstart} with filled circles.
The continuous portions of the fast oscillations tend to start 
later in the burst
(with only two exceptions) and to drift by a smaller absolute frequency 
range.  We believe that this is 
due to the fact that the faster oscillations tend to occur in bursts with 
radius expansion (Muno et

\begin{inlinefigure}
\centerline{\epsfig{file=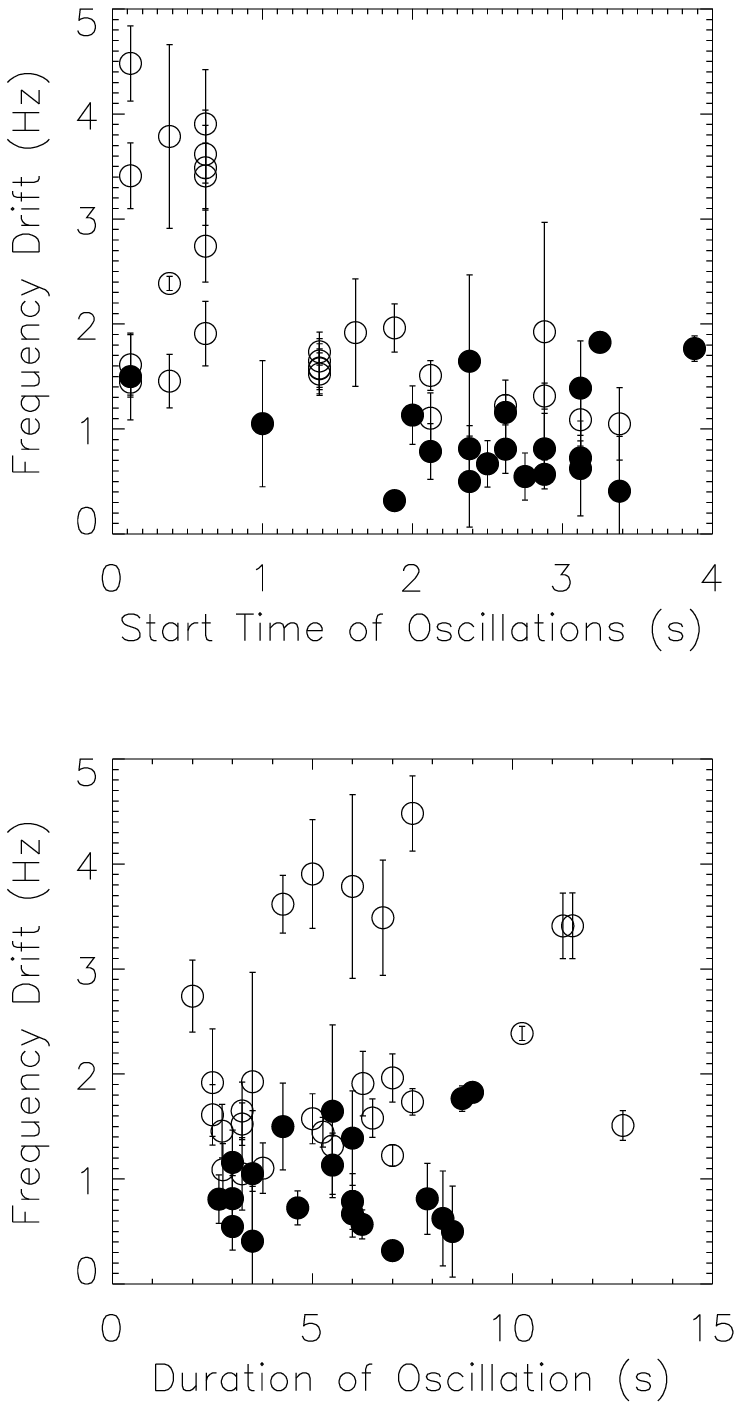,width=0.65\linewidth}}
\caption{The total frequency drift in the continuous portion of the 
 oscillation trains plotted as a function of the ({\it top panel}) the 
time at which 
the continuous oscillation train began, and ({\it bottom panel}) the 
duration of the continuous train. The total frequency drift is largest 
when the 
oscillations are first observed at the start of the bursts, but is not 
correlated with the duration of the oscillations. The solid circles are 
data from fast ($\sim$600~Hz) oscillations, while the open circles are 
from slow ($\sim$300~Hz).
\label{tstart}}
\end{inlinefigure}

\noindent
al.\ 2001),
which prevents the oscillations from 
being observed earlier in the bursts (see Section~3.1). Indeed, the 
fast oscillators in Figure~\ref{hist} do not in general have a narrower 
distribution of observed frequencies when considering the non-continuous 
portions of the oscillations as well. However, we do confirm that the 
fractional frequency drift is a factor of two smaller in the 
$\sim$600~Hz oscillators
than in the $\sim$300~Hz ones, but only because the frequencies $\nu$ are
larger.

If the frequencies of the oscillations are associated with underlying stable clocks,
then the fact that they evolve in frequency implies that the phases of the
oscillations drift by several cycles with respect to those clocks. If we take 
the maximum observed frequency ($\nu_{\rm max}$) to represent the stable clock 
(see Section~3.7), then we can define a phase loss by 
\begin{equation}
\Delta\phi_{\rm loss} = \nu_{\rm max}t - \int \nu(t) dt,
\end{equation}
where $\nu(t)$ is the derivative of our phase model.
We find that oscillations typically lose between 1--5 cycles, depending on the magnitude 
of the frequency drift, the duration of the oscillations, and the time scale for the 
frequency evolution. One exceptional oscillation, illustrated in Figure~\ref{consev}b, 
drifts by more than 17 cycles with respect to $\nu_{\rm max}$. This oscillation is 
unusual for its combination of a long duration (10~s), a large frequency
change ($\Delta\nu = 4.1$~Hz), and a slow drift time scale ($\tau = 5.5$~s).

\subsection{``Asymptotic'' Frequencies of Burst Oscillations}

Aside from the exceptions noted above, the fundamental features 
that led to the adoption of an exponential model for the frequency 
evolution by \citet{sm99} are present in about $70\%$ of burst oscillations:
the oscillations increase in frequency when they are first observed, but
appear to reach stable frequencies on the time scale of seconds. 
Several authors have noted that the asymptotic
frequencies of burst oscillations detected several years apart are remarkably
similar \citep{str98, sm99, mun00}.  
Therefore, we measured the maximum frequencies that the oscillations 
reached according to our best fit models, in order to determine whether
the apparent saturation frequencies of the oscillations are stable.
The 1-$\sigma$ uncertainties on the maximum 
frequency were found by a search through $\chi^2$ space \citep[e.g][]{lmb76}.
In those instances where the best available fit was statistically 
unacceptable, we inflated the uncertainties on the phase measurements to for 
the reduced $\chi^2$ to equal 1 before the search in $\chi^2$ space. The 
maximum frequency was 
constrained to occur after a minimum, as in several cases the absolute highest
frequency was observed at the start of the oscillation train 
\citep[e.g][and Figure~3a]{mil00}. Our measurements are consistent with 
those found 
using the $Z^2$ statistic by \citet{str98}, \citet{sm99}, and \citet{gil02}.
We compare our results for \sixb\ with those of \citet{gil02} in Section~4.2

\begin{figure*}[t]
\centerline{\epsfig{file=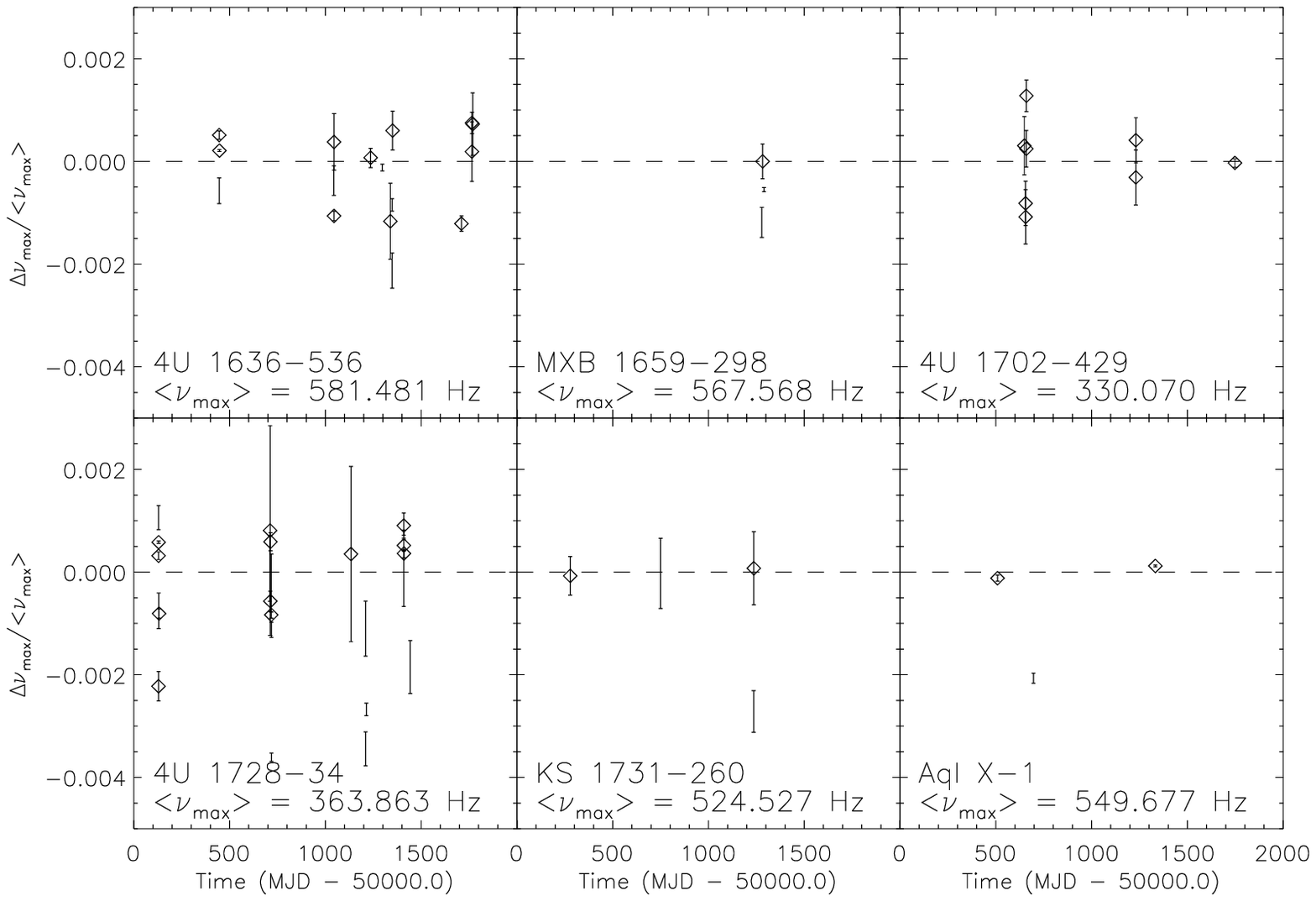,width=0.85\linewidth}}
\caption{The maximum frequencies from Table~1 plotted as the fractional
deviation from the mean as a function of time. Measurements from 
oscillation
trains which appear to saturate, and are therefore more likely to be 
indicative of an ``asymptotic'' frequency, are marked with diamonds, 
while those which do not are simply reported with error bars.
\label{dispersion}}
\end{figure*}

In Figure~\ref{dispersion}, we plot how the maximum frequencies vary with time.
To indicate those oscillations where the frequency has apparently saturated, 
we have indicated which ones required at least an exponential or a 
3rd order polynomial phase model by diamonds. 
The measurements taken from first or second order polynomial models 
are plotted only with error bars.
We have calculated the fractional standard deviation in the maximum 
frequencies, $\sigma_\nu/\langle\nu_{\rm max}\rangle$, 
for those sources with more than two values, 
and listed them in Table~\ref{tab:disp}. For \aqlxone\ and \ksxrb, the 
frequency evolution appeared to saturate during two bursts each, so we 
report the fractional difference between the measurements. The deviations 
in these frequencies  
$\sigma_\nu/\langle\nu_{\rm max}\rangle$ are quite small, less 
than one part in 1000.

\section{Discussion}

The frequency evolution of burst oscillations can be described by smooth 
phase models in all but 30\% of the pulsations that we have studied
(Section~3.5; see also Strohmayer 2001). The discrepant cases appear as 
phase jumps in Figure~\ref{failed}.
If the oscillations result from brightness asymmetries on the surface
of the neutron star, it appears that some of the following occur: 
({\it i}) there are multiple 
anisotropies or modes that propagate with slightly different velocities
relative to the surface, which results in two simultaneous signals that
interfere with each other (see also Section~3.2), ({\it ii}) the
velocity of the pattern changes suddenly on time scales of 0.25 s, or 
({\it iii}) the anisotropies change their longitude in less than a 
tenth of a second, resulting in sudden phase jumps of about 0.1 cycles.

On time scales of years, the oscillations are stable to a few parts in 
1000. This is consistent with the Doppler 

\noindent
shifts that are expected from the binary orbit.
Since this raises the possibility of measuring a mass function for the
companion star, we now examine this frequency dispersion in detail.

\subsection{The Stability of the Maximum Frequencies}

The change in the apparent spin frequency due to the orbital motion 
of the neutron star is given by 
\begin{equation}\label{eq:dopp}
 {{\Delta\nu_{\rm max}} \over {\langle\nu_{\rm max}\rangle}} 
  = 2.0 \times 10^{-3} {{M_c \sin{i}} \over
{P_{hr}^{1/3} (M_x + M_c)^{2/3}}},
\end{equation}
where $M_x$ and $M_c$ are the mass 
of the neutron star and companion respectively in solar units,
$P_{hr}$ is the binary orbital period in hours, and $i$ is the inclination
of the orbital plane normal to the line of sight. Assuming a star with 
solar density that over-fills it
Roche lobe, the mass and orbital period are related by 
$M_c \approx 0.11 P_{\rm hr}$ (Frank, King, \& Raine 1995). 
For $P_{hr} = 5$, $\sin i=0.87$, $M_x=1.4$, and $M_c=0.5$,
the half-amplitude of the Doppler variation would be 
$\Delta\nu_{\rm max}/\langle\nu_{\rm max}\rangle = 4\times10^{-4}$.
This is a factor of 2--5 smaller than the dispersions from 
\sixb, \sevenb, and \slowb\ (Figure~\ref{dispersion}), but it is 
of the right order of magnitude.

Only three oscillation sources have known orbital ephemerides: 
\mxbecl\ \citep{wsb00}, \aqlxone\ \citep{wry00}, and \sixb\ \citep{gil02}.
We did not measure enough oscillations from \mxbecl\ and \aqlxone\ to 
constrain the parameters of the binary orbit.
We have measured eleven oscillations that appear to saturate from
\sixb\ (the diamonds in Figure~\ref{dispersion}), so we have attempted
to fit a sinusoid to the maximum frequencies as a function of orbital
phase (Figure~\ref{phase}). The best-fit sinusoid, allowing the 
reference phase to 
vary, has a $\chi^2$ of 221 for 8 degrees of 
freedom. Therefore, sinusoidal orbital 
modulations cannot cause all of the observed dispersion. In Figure~\ref{phase}
we plot a sinusoid with an amplitude equal to the 
standard deviation of the points fit, 
$\sigma_\nu/\langle\nu_{\rm max}\rangle = 7.7\times10^{-4}$.
The disagreement between the data and the sinusoid is primarily due
to three oscillations with low maximum frequencies. Two of these 
lasted for more than 5 s, one of which is displayed in the right panel of
Figure~\ref{consev} (1999 June 10). We will discuss this further in 
the next section.

There are no other effects that ordinarily change the apparent spin 
frequencies of neutron stars (i.e. pulsars) that can produce the
dispersion in Figures~\ref{dispersion} and \ref{phase}. Accretion torques 
can alternately 
quicken or slow the rotation of the neutron star, and would provide
the next largest effect \citep{fkr95}. The torque on a neutron star
can be estimated by assuming that its magnetic field disrupts the disk 
at some fiducial radius $R_M$ and removes all of its angular momentum. At 
$R_M$, we have $I\dot{\Omega} = \dot{M}(GMR_M)^{1/2}$. 
We take $R_M$ to be the co-rotation radius, where 

\begin{inlinefigure}
\centerline{\epsfig{file=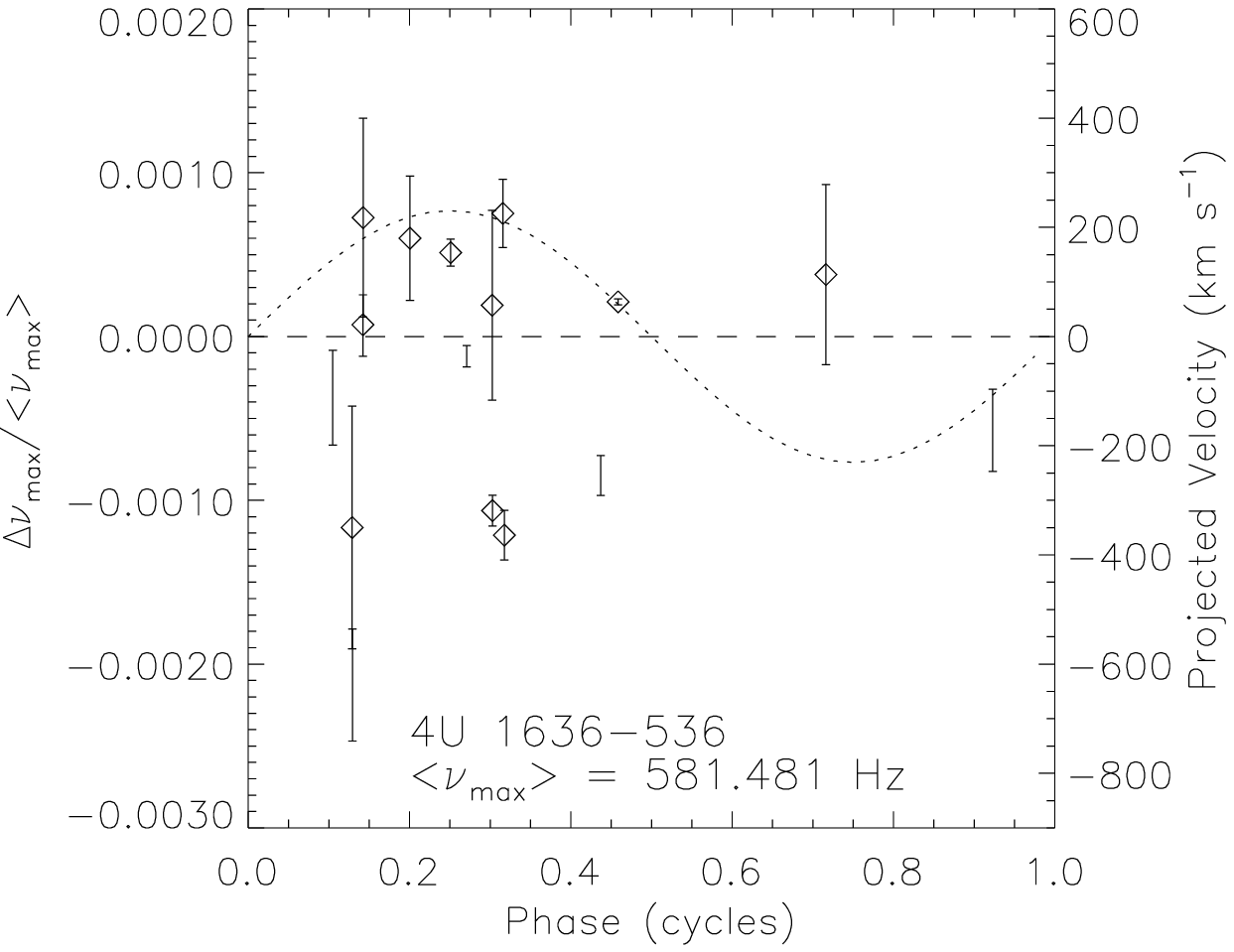,width=0.9\linewidth}}
\caption{The fractional deviation of the maximum frequencies from Table~1 
are plotted as a function of orbital phase for \sixb\ 
(see also Figure~6). 
The dotted line indicates the expected orbital modulation using the 
ephemeris of \citet{gil02}.
The dispersion in asymptotic frequencies can not be explained by 
orbital motion
alone. \label{phase}}
\end{inlinefigure}

\noindent
$2\pi\nu_{\rm kHz} = (GM/R_M^3)^{1/2}$, so that
\begin{equation}\label{eq:torque}
\dot{\nu} = 7\times10^{-5} \dot{M} R_{6} M_x^{2/3} I_{45}^{-1} 
\nu_{\rm kHz}^{-1/3}
{\rm Hz~yr}^{-1}
\end{equation}
where $\dot{M}$ is the accretion rate in Eddington units, 
$R_{6}$ is the radius of the neutron star in units of 10~km, 
$M_x$ is its mass in solar units, $I_{45}$ 
is its moment of inertia in units of $10^{45}$ g cm$^2$, 
and $\nu_{\rm kHz}$ is the stellar spin frequency in kHz. 
The values are scaled to those expected for a burster. For a spin 
frequency of 200~Hz, Equation~\ref{eq:torque} corresponds to a 
$\Delta\nu_{\rm max}/\langle\nu_{\rm max}\rangle \approx 6 \times 10^{-7}$ 
in one year. Clearly, the 
observed dispersion in the maximum frequencies on time scales of a month  
(Figure~\ref{dispersion}) is much larger than can be explained by accretion 
torques (Equation~\ref{eq:torque}).

\subsection{Frequency Dispersion in \sixb}

\citet{gil02} recently examined a larger sample of oscillations
from \sixb, and found that there is a subset
of the burst oscillations whose maximum frequencies 
are distributed as a Gaussian about 581.64 Hz with 
a width of 0.08 Hz. We cannot independently test this, as the uncertainties 
that we report on the maximum frequencies are up to a factor of 5 larger than
those in \citet{gil02}. This is because the phase-connection technique we 
used to measure the maximum frequencies accounts for
the frequency evolution of the oscillations, while the dynamic $Z^2$ 
technique \citet{gil02} used assumes that the frequency 
is constant during each 2 s interval that they test. In our analysis, 
the maximum frequency is usually observed at the end of the oscillation 
train where the data can tolerate relatively large variations in the values 
of the polynomial models.
\footnote{On the other hand, we may be 
over-estimating our uncertainties when we model the oscillations with 
higher-order polynomials, which tend to have large derivatives at their 
endpoints. A more physical model might assume that the frequency change in 
the oscillations is much smaller at the endpoints. However, the oscillation 
that appears 4~Hz from the main signal at the end of the burst on 
1999 Apr 14 from \mxbecl\ \citep{wsf00} makes us reluctant to apply this 
assumption in our analysis, and leads us to assume larger uncertainties.}
In the analysis of \citet{gil02}, if significant frequency evolution occurs
during a 2 s interval, they measure the mean frequency in that 
interval, but do not consider how much larger the frequency could be 
at the end of the interval.
As a result the magnitude of our uncertainties, 0.05--0.4~Hz, 
are often larger than the 
width of the Gaussian of \citet{gil02}.

If we only include those oscillations that are part of the Gaussian in 
\citet{gil02}, we find that the data are consistent with no modulation. 
Fitting a sinusoid to the data, we derive a fractional amplitude of
$(5.0\pm2.1)\times10^{-4}$ ($\chi^2=6.7$ for 5 
degrees of freedom). This is equivalent to a Doppler 
shift of 150$\pm$63 km s$^{-1}$, 
which is significantly larger than the 90\% upper 
limit to the velocity from Giles 
et al. (2002; 55 km s$^{-1}$). The larger 
uncertainties that we assumed result in larger upper limits on the
velocity. These systematic uncertainties also 
present an additional hurdle in measuring the orbital motion of the neutron 
star. 

\subsection{Models for the Frequency Evolution}

Three models have been proposed to explain the frequency evolution of 
the burst oscillations, all of which assume that anisotropies develop 
in the surface brightness of the neutron star
\citep{str97,cb00,hey02,slu02}. The simplest model was proposed 
by \citet{str97}, who suggested that the 
oscillations originate from hot regions on a burning layer that 
expands and decouples from the neutron star at the start of the burst. 
The oscillations are observed when the burning layer begins to 
contract, causing them to increase in frequency as the layer 
re-couples to the neutron star. However, recent calculations assuming that 
the burning layer is rigidly rotating predict a frequency drift a factor of 
three smaller than that observed \citep{cum02}. Moreover, \citet{cb00} 
predicted that there should be a fractional dispersion of at most 
$\Delta\nu/\nu \approx 10^{-4}$ in the observed asymptotic frequencies, 
if the height of burning layer is smaller when it re-couples to the star
than when it decouples. This is a factor of 10 smaller than the residual 
dispersion from \sixb\ in Figure~\ref{phase}.

As a result, it is important to consider other models for 
the frequency evolution of burst oscillations. 
\citet{slu02} have suggested that a brightness contrast could result
from a hydrodynamical instability in a geostrophic flow, 
similar to Jupiter's Great Red Spot, while \citet{hey02} has proposed 
that Rossby waves propagating retrograde on the
neutron star could produce traveling brightness patterns on the stellar
surface. The frequency evolution in both models is caused by changes in 
the velocity at which the pattern propagates around the star. The velocity
in turn is sensitive to the temperature, density, and vertical structure of 
the surface layers. Thus, variations in the properties of the surface
layers near the end of a burst can naturally explain the observed 
dispersion in the asymptotic frequencies of the oscillations 
(Figures~\ref{dispersion} and \ref{phase}).
On short time scales, instabilities in either mechanism conceivably 
could cause the apparent phase jumps in Figure~\ref{failed}. Moreover,
\citet{hey02} also has predicted that Rossby waves with different numbers 
of radial nodes could produce simultaneous signals at multiple frequencies,
as are seen in rare cases (Section~3.2; compare Section~3.5). These models 
are promising, although how these mechanisms produce brightness variations on 
the stellar surface remains to be studied in detail.

\section{Conclusions}

We have examined 68 burst oscillations observed in 159 bursts from 8 
different sources, and have modeled the frequency evolution of 59 
oscillations from 6 sources (Table~\ref{bestmod}). The most notable 
result is that there appears to be slight instability in the mechanism 
generating the oscillations on time scales of both seconds and years. 
About 30\% of the oscillation trains do not exhibit smooth frequency 
evolution, as both low order polynomials and exponential
models are statistically inconsistent with the data (Figure~\ref{failed}). 
The possible explanations are that ({\it i}) two signals may be present 
simultaneously, invalidating our assumption that there is a single signal 
in our analysis, ({\it ii}) there are discrete phase jumps that occur on time 
scales less than a second, and ({\it iii}) the frequencies of the oscillations
shift dramatically on time scales of 0.25 s.
On longer time scales, the maximum frequencies 
observed during these oscillation trains exhibit a fractional dispersion of 
$\Delta\nu_{\rm max}/\langle\nu_{\rm max}\rangle \la 4\times 10^{-3}$
(Figure~\ref{dispersion}).
For 5 of the sources studied, this dispersion could be consistent with 
Doppler shifts of signals originating from rotating neutron stars as they 
orbit the centers of mass of their binaries. However, the dispersion in the 
maximum frequencies in \sixb\ is uncorrelated with its known orbital period 
(Figure~\ref{phase}). 

We now are able to accurately characterize the general properties 
of the frequency evolution in our large sample of bursts:

\bigskip
1) The frequency of the oscillations drifts by up to 1.2\% 
(Figure~\ref{fig:typical} and \ref{hist}), but generally
reaches a stable value on the time scales of seconds. 

2) Both spin-down in the oscillation train and obvious examples of 
simultaneous signals
separated by around 1 Hz are relatively rare, occurring in 3 and 2 bursts
out of 65 respectively (Figure~\ref{sdown} and \ref{fig:twofreq}). 

3) The amounts of the frequency drifts appear larger 
when oscillations are observed earlier in the bursts (Figure~\ref{tstart}), but 
are uncorrelated with the durations of the oscillations. This implies that the starts of 
the bursts set in motion the eventual changes in the oscillation frequencies, as opposed to 
some mechanism that operates when the oscillations themselves appear.

4) Photospheric radius expansion appears to temporarily interrupt 
oscillations. All of the 16 oscillation trains that appear 
continuously from the rise to the tail of the burst fail to exhibit 
radius expansion. 
When radius expansion does occur, oscillations are often seen in the 
rise and/or the tail of the burst, but rarely during the radius expansion
(Figure~\ref{fig:pre}). 

5) Aside from the above points, the time scale and magnitude of the 
frequency evolution does not appear correlated with other properties of
the burst, such as its decay time scale, peak flux, or fluence.

6) In about 60\% of
the oscillation trains, the frequency evolution displays inflections that 
cannot be described by an exponential model. 
Two illustrative examples are shown in Figure~\ref{consev}.
Thus, in addition to the upward frequency drift, the frequency tends to 
wander by on order 0.1~Hz in a random manner.
\bigskip

Several models have been proposed in which the oscillations originate from 
an anisotropy in the emission from the neutron star's surface. Each 
models the frequency evolution as originating from different mechanisms
\citep{str97,cum02,hey02,slu02}. The apparent lack of stability in the 
underlying clock producing these oscillations favors the models in 
which the brightness asymmetry originates from hydrodynamic 
instabilities \citep{slu02}
or modes excited in the neutron star ocean \citep{hey02}.
Future studies of the amplitude, harmonic structure, energy spectrum, and 
phase lags of the oscillations will further constrain the location of 
the hot spot and probe the atmosphere of the neutron star.

\acknowledgements{We thank Lars Bildsten and Fred Lamb for useful comments,
and the referee for careful reading of this manuscript. We also thank Pavlin
Savov for developing the burst detection software, and Derek Fox 
for developing much of the phase connection software.
This work was supported by NASA, under contract NAS 5-30612 and grant 
NAG 5-9184.}

\begin{deluxetable}{lcccc}
\tablecolumns{5}
\tablewidth{0pc}
\tablecaption{Summary of Burst Oscillations\label{sources}}
\tablehead{
 \colhead{} & \colhead{$\nu_{\rm burst}$} & 
   \multicolumn{3}{c}{Number of} \\
 \colhead{Source} & \colhead{(Hz)} & \colhead{Bursts} & 
   \colhead{Osc.} & \colhead{Models\tablenotemark{a}}
}
\startdata
\novab\ & 620 & 6 & 1 & 0 \\ 
\sixb\ & 581 & 24 & 17 & 17 \\
\mxbecl\ & 567 & 15 & 5 & 3\\
\aqlxone & 549 & 17 & 3 & 3 \\
\ksxrb\ & 524 & 13 & 5 & 4\\ 
\slowb\ & 363 & 66 & 27 & 24 \\
\sevenb\ & 329 & 11 & 8 & 8 \\
\degdip\ & 270 & 7 & 1 & 0 \\
\enddata
\tablenotetext{a}{The number of oscillations modeled using the phase 
connection technique.}
\end{deluxetable}

\begin{deluxetable}{lc}
\tablecolumns{2}
\tablewidth{0pc}
\tablecaption{Models for the Phase Evolution of Burst Oscillations
\label{bestmod}}
\tablehead{ \colhead{Best} & \colhead{Number}\\
 \colhead{Model} & \colhead{of Oscillations}}
\startdata
1st Order & 9\\
2nd Order & 13\\
3rd Order & 18\\
4th Order & 3\\
5th Order & 10\\
Exponential & 6\\
\enddata
\end{deluxetable}

\begin{deluxetable}{lc}
\tablecolumns{2}
\tablewidth{0pc}
\tablecaption{Dispersion in Maximal Frequencies of Burst Oscillations
\label{tab:disp}}
\tablehead{ \colhead{Source} & 
\colhead{$\sigma_{\nu}/\langle\nu_{\rm max}\rangle$} }
\startdata
\sixb\ & 7.7$\times10^{-4}$ \\
\sevenb\ & 7.5$\times10^{-4}$ \\
\slowb\ & 9.2$\times10^{-4}$ \\
\ksxrb\tablenotemark{a} & 7.3$\times10^{-5}$\\
\aqlxone\tablenotemark{a} & 1.2$\times10^{-4}$\\
\enddata
\tablecomments{$\sigma_{\nu}$ is defined as the standard deviation 
of the 
maximum frequencies measured in oscillations that appear to saturate, 
unless otherwise indicated.}
\tablenotetext{a}{Only two maximum frequencies were measured, so 
$\sigma_\nu$
is the difference between the values.}
\end{deluxetable}

\end{document}